# A Feasibility Study on Deep Learning–Based Individualized 3D Dose Distribution Prediction[1]

Jianhui Ma[1,2], Dan Nguyen[2], Ti Bai[2], Michael Folkerts[2], Xun Jia[2], Weiguo Lu[2], Linghong Zhou[1, a)] and Steve Jiang[2, a)]

[1] *School of Biomedical Engineering, Southern Medical University, Guangzhou, Guangdong 510515, China*
[2] *Medical Artificial Intelligence and Automation (MAIA) Laboratory, Department of Radiation Oncology, University of Texas Southwestern Medical Center, Dallas, TX 75235, USA*

a) Co-corresponding authors. Emails: smart@smu.edu.cn and Steve.Jiang@utsouthwestern.edu

**Abstract**

**Purpose**: Radiation therapy treatment planning is a trial-and-error, often time-consuming process. An approximately optimal dose distribution corresponding to a specific patient's anatomy can be predicted by using pre-trained deep learning (DL) models. However, dose distributions are often optimized based not only on patient-specific anatomy but also on physicians' preferred trade-offs between planning target volume (PTV) coverage and organ at risk (OAR) sparing or among different OARs. Therefore, it is desirable to allow physicians to fine-tune the dose distribution predicted based on patient anatomy. In this work, we developed a DL model to predict the individualized 3D dose distributions by using not only the patient's anatomy but also the desired PTV/OAR trade-offs, as represented by a dose volume histogram (DVH), as inputs.

**Methods:** In this work, we developed a modified U-Net network to predict the 3D dose distribution by using patient PTV/OAR masks and the desired DVH as inputs. The desired DVH, fine-tuned by physicians from the initially predicted DVH, is first projected onto the Pareto surface, then converted into a vector, and then concatenated with feature maps encoded from the PTV/OAR masks. The network output for training is the dose distribution corresponding to the Pareto optimal DVH. The training/validation datasets contain 77 prostate cancer patients, and the testing dataset has 20 patients.

**Results:** The trained model can predict a 3D dose distribution that is approximately Pareto optimal while having the DVH closest to the input desired DVH. We calculated the difference between the predicted dose distribution and the optimized dose distribution that has a DVH closest to the desired one for the PTV and for all OARs as a quantitative evaluation. The largest average error in mean dose was about 1.6% of the prescription dose, and the largest average error in the maximum dose was about 1.8% of the prescription dose.

**Conclusions:** In this feasibility study, we have developed a 3D U-Net model with the patient's anatomy and the desired DVH curves as inputs to predict an individualized 3D dose distribution that is approximately Pareto optimal while having the DVH closest to the desired one. The predicted dose distributions can be used as references for dosimetrists and physicians to rapidly develop a clinically acceptable treatment plan.

## 1. Introduction

With the rapid development of external beam radiotherapy, the treatment planning procedure has become increasingly complicated for many tumor sites. In the current treatment planning workflow, a treatment planner works towards a good quality plan in a trial-and-error fashion by using a commercial treatment planning system. In the meantime, many rounds of consultation between planner and physician are often needed to reach a plan that meets the physician's satisfaction for a particular patient, mainly because medicine, to some degree, is still an art, and a physician's preferences for a particular patient cannot be quantified and precisely conveyed to the planner. Consequently, planning can take up to a week for complex cases, and plan quality may be poor and can vary significantly according to the varying levels of the physician's and the planner's skills, the quality of their communication, etc.[1,2]

---

[1] This work was initially presented at the Annual Conference of American Association of Physicists in Medicine in Nashville, TN, 2018.

To rapidly produce a treatment plan with consistent plan quality, knowledge-based planning (KBP)[3-5] was developed to build a heuristic correlation between patient anatomy and the best achievable dose volume histogram (DVH) based on the treatment plans of previously treated patients. This provides useful clues as to how a plan of acceptable quality should look for any particular patient by looking at common features among similar patients treated in the past. Yet such methods heavily rely on the selection of handcrafted features to train the linear regression model. These handcrafted features, such as distance-to-target histograms, overlapping volume histograms, and organ shapes, are oversimplified representations of patient anatomy (e.g., they only focus on the pair-wise relationship between the target and one particular organ at a time) and thus cannot precisely predict the treatment plan with the best achievable quality. Moreover, current KBP approaches can only predict DVH curves, not 3D dose distributions; these DVH curves provide an incomplete description of the plan quality.

Over the past couple of years, deep learning (DL) has been used for predicting dose distribution. Nguyen *et al.* first explored the feasibility of predicting dose distributions from organ contours by utilizing a modified U-Net model for prostate cancer and then extended their work to more complicated head and neck cancer cases.[6-8] Their model can automatically extract critical features from a patient's anatomy without any handcrafted parameters to precisely predict the dose distributions. Barragán-Montero *et al.*[9] extended this work to develop a more general model that considers variable beam configurations in addition to patient anatomy to predict dose distributions for lung intensity-modulated radiation therapy (IMRT), thus indicating a potentially easier clinical implementation with no need to train specific models for different beam settings. Similar ideas have been implemented by different research groups and applied to various clinical scenarios.[10-13]

In theory, there is an unlimited number of optimal treatment plans corresponding to a particular patient anatomy, due to the different trade-offs between the coverage of the planning target volume (PTV) and the sparing of various organs at risk (OARs). These optimal plans constitute the so-called Pareto surface. Therefore, dose prediction should also consider, in addition to patient anatomy, the trade-off preferred by the attending physician for a particular patient. There are also other clinical factors that must be considered in the planning process. Therefore, when a dose prediction model is used to generate a dose distribution and the corresponding DVH based a patient's anatomy, the physician may often need to tune the results based on some specific considerations for the particular patient to produce the directive for the planner to follow.

Nguyen *et al.*[14] trained a neural network for generating Pareto optimal dose distributions by proposing a differentiable loss function based on the DVH and combining it with an adversarial loss to train deep neural networks. Although this model can predict Pareto optimal dose distributions, it requires structure prioritization weights as part of the input. These weights are essentially abstract concepts related to, but not equal to, clinically relevant metrics and values, which makes the model less accessible to physicians.

In this work, we propose to develop and test the feasibility of another DL model to predict the optimal dose distribution from the patient anatomy and PTV/OARs trade-offs represented by a set of desired DVH curves. We envision the following clinical workflow, illustrated in Figure 1: 1) a conventional anatomy-based DL model will be used to predict a dose distribution first, 2) if the predicted dose distribution is not the desired one, the physician will tune the DVH of the predicted dose distribution to reflect the individualized planning goals for the particular patient, and 3) the proposed DL model will be used to predict a refined dose distribution by using the patient anatomy and tuned DVH as inputs. Steps 2) and 3) can be repeated until the desired dose distribution and DVH are achieved.

The remainder of this paper is organized as follows. Section 2 first introduces the framework of our method and the network architecture. Then, it describes the dataset and the training configuration. The performance of our proposed method is presented in Section 3. Section 4 discusses and summarizes the strengths and weaknesses of the proposed model.

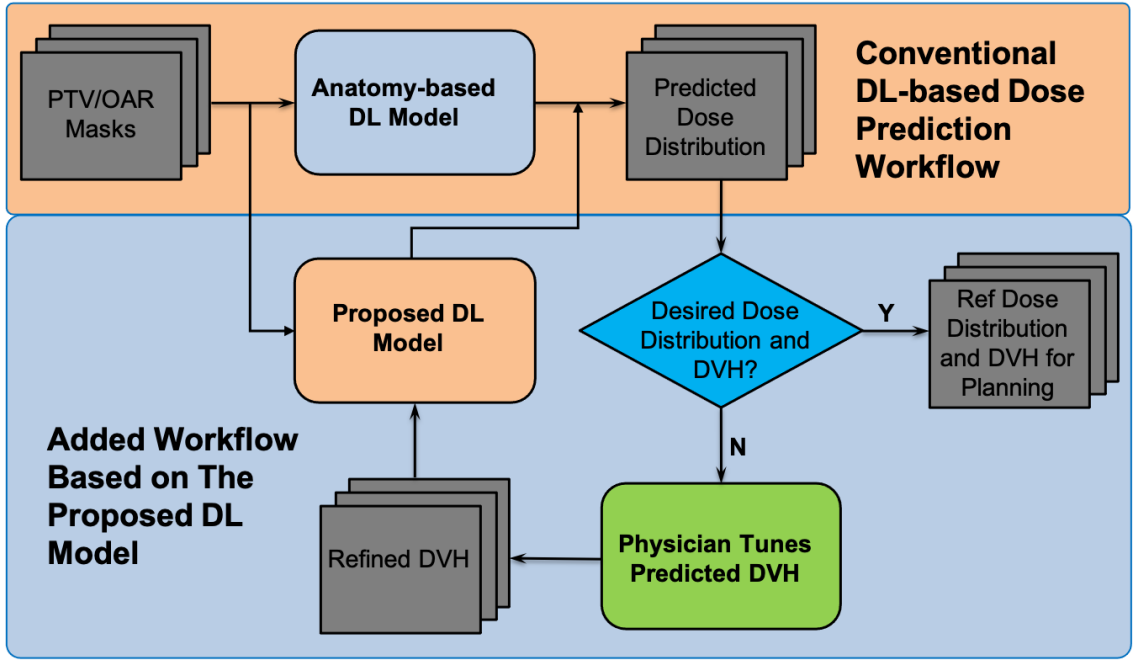

Figure 1. The envisioned clinical workflow based on the proposed DL model.

## 2. Methods and Materials

We first mathematically formulate our task. Basically, given patient CT anatomy $X$, we would like to predict the associated 3D dose distribution $Y$, which can be achieved by maximizing the following conditional probability:

$$max\ P(Y|X) \qquad (1)$$

As mentioned above, in practice, problem (1) is underdetermined and therefore has multiple optimal solutions corresponding to different trade-offs among PTV coverage and OAR sparing. The DL models previously developed for dose prediction generate the optimal dose distributions by averaging all PTV/OAR trade-offs presented in the training dataset. In this work, we introduce an extra condition—i.e., a set of DVH curves that can represent the desired trade-offs—to constrain the solution space to one specific point on the Pareto surface. In other words, the objective function can now be expressed as:

$$max\ P\ (Y^D|X, D) \qquad (2)$$

where $D$ represents the desired DVH curves and $Y^D$ is a point on the Pareto surface corresponding to the sample pair $(X, D)$.

Now the problem (2) is well-defined and can be solved in a learning-and-prediction fashion. To be specific, suppose we have a training dataset $\{X_i, D_i^j, Y_i^j | i \in 1, \cdots, M, j \in 1, \cdots, N\}$ where $i$ indexes the patients and $j$ indexes the sampling point on the Pareto surface. We assume that there are $M$ patients in the dataset and $N$ data points on the Pareto surface for each patient, meaning we have $N$ pairs of Pareto optimal DVH curves and dose distributions. Then, problem (2) can be solved based on the above training dataset by training a convolutional neural network (CNN) $\phi_w(X, D)$ parameterized by $w$. If one uses mean squared error (MSE) as the cost function, the network parameter $w$ can be calculated by solving the following specialized cost function:

$$\bar{w} = arg\ \min_w \sum_{i=1}^{M}\sum_{j=1}^{N} \left\|\phi_w(X_i, D_i^j) - Y_i^j\right\|_2^2 \qquad (3)$$

Once the network $\phi_{\bar{w}}(X, D)$ is trained, in theory, it can produce the optimal dose distribution if it is fed as inputs the patient anatomy $X$ and the DVH vectors $D$ that describe the desired trade-off between PTV coverage and OAR sparing.

One may notice that, based on our envisioned clinical workflow, we want to predict a dose distribution that is approximately Pareto optimal, although the physician-tuned DVH $\mathcal{D}_i$ is unlikely to correspond to a Pareto optimal dose distribution. Therefore, when training the model, we need to first project $\mathcal{D}_i$ onto the Pareto surface of the patient $i$ to find the nearest Pareto optimal DVH $D_i^j$ and the corresponding dose distribution $Y_i^j (j \in 1, \cdots, N)$ from the training dataset (Fig. 2). $Y_i^j$ is termed "Pareto optimal dose distribution" in Figure 2 and used as the ground truth for model training. In this work, we use $l_1$-norm to conduct the DVH projection operation, i.e., we solve the following problem to select the Pareto optimal DVH $D_i^j$ from the training dataset:

$$S_i = arg \min_{S_i} ||\boldsymbol{D}_i S_i - \mathcal{D}_i||_1^1 \quad (4)$$

where each column of the matrix $\boldsymbol{D}_i$ represents one point on the Pareto surface, which corresponds to an optimal DVH for patient $i$, i.e., $\boldsymbol{D}_i = |D_i^0, D_i^1, \cdots, D_i^j, \cdots, D_i^N|$. $S_i$ is an N-length one-hot vector, which is used to indicate the selection state.

Let us assume that one can generate $K$ desired DVHs for each training patient. We denote the $k$th desired DVH of patient $i$ as $\mathcal{D}_i^k$, and the corresponding one-hot vector can be calculated as $S_i^k$ according to equation (4). Then, the associated dose distribution can be calculated as $\mathcal{Y}_i^k = \boldsymbol{Y}_i S_i^k$, where $\boldsymbol{Y}_i = |Y_i^0, Y_i^1, \cdots, Y_i^j, \cdots, Y_i^N|$.

By using this projection operation, we now convert our original training dataset with $N$ pairs of Pareto optimal DVH curves and dose distributions, $\{X_i, D_i^j, Y_i^j | i \in 1, \cdots, M; j \in 1, \cdots, N\}$, into an augmented training dataset as $\{X_i, \mathcal{D}_i^k, \mathcal{Y}_i^k | i \in 1, \cdots, M; k \in 1, \cdots, K\}$. Here, the data points $\{X_i, D_i^j, Y_i^j\}$ in the original training dataset are generated and saved before model training, while the desired DVH $\mathcal{D}_i^k$ and the association relationship $\{\mathcal{D}_i^k, S_i^k\}$, and hence $\{\mathcal{D}_i^k, \mathcal{Y}_i^k\}$ in the augmented training dataset, can be calculated in an on-the-fly fashion during the training phase by solving problem (4).

In this training strategy, the cost function can be expressed as:

$$\overline{w} = arg \min_{w} \sum_{i=1}^{M} \sum_{k=1}^{K} ||\phi_w(X_i, \mathcal{D}_i^k) - \mathcal{Y}_i^k||_2^2 \quad (5)$$

For the testing phase, one just needs to feed the network with the patient contours and a desired (unlikely Pareto optimal) DVH to predict an approximately Pareto optimal dose distribution prediction with a DVH close to the desired one, without needing the DVH projection operation, i.e., $\overline{Y}_i = \phi_w(X_i, \mathcal{D}_i^k)$.

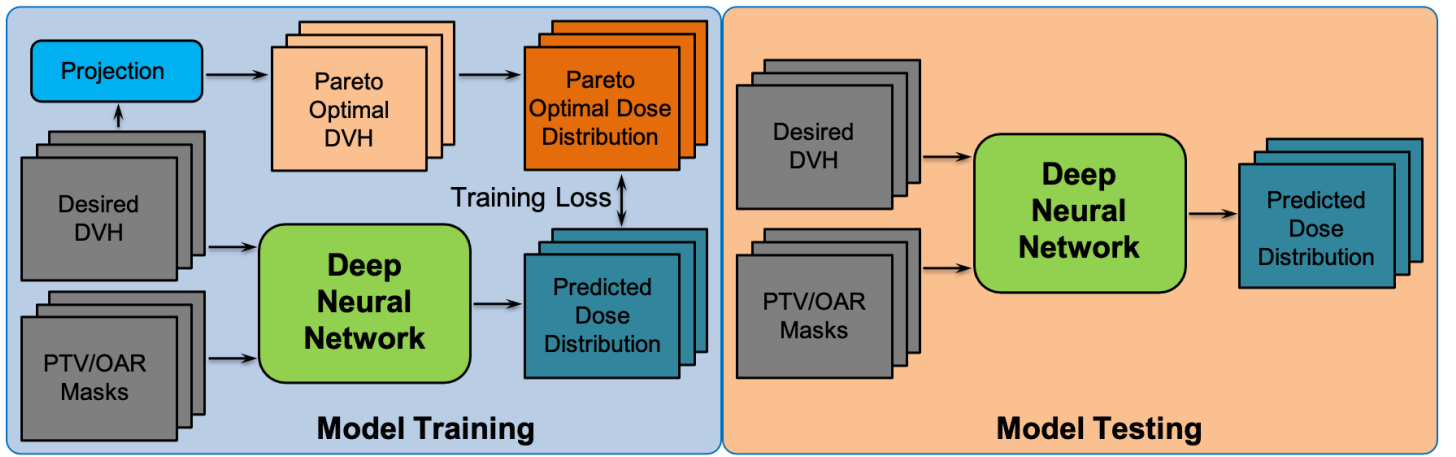

Figure 2. The workflow of the proposed method for model training and testing.

*2.1 Network Architecture*

Figure 3 details the architecture design for the deep neural network employed in Figure 2. In this work, we use a modified 3D U-Net architecture with an encoder (left half) and a decoder (right half) as the architecture. In more detail, the encoder first extracts the features from the patient anatomy input, which is a multi-channel representation, each channel representing the contour of an OAR or the PTV. Then, these patient anatomy features are concatenated with the desired DVH before they are fed into the decoder for individualized dose prediction. It should be noted that, in this work, the DVH is a vector representation, i.e., the dose volume histograms of each OAR and PTV are stacked into a 1D vector.

For the inner architecture, in the encoder part, we use consecutive convolutional layers to extract the features (Fig. 3). Each layer contains three operators: a convolutional operator, a batch normalization (BN) operator, and a rectified linear unit (ReLU). All the convolutional operators have a kernel size of 3 × 3 × 3, except the last two layers, which use a kernel size of 1 × 1 × 1 instead because of the limitation of the feature map size. Zero padding is applied to keep the feature size invariant during the convolution process. Six max-pooling operations with a 2 × 2 × 2 pooling size are employed to reduce the input size from 128 × 128 × 64 to 2 × 2 × 1, then one max-pooling with a 2 × 2 × 1 pooling size is utilized to obtain the feature maps with a size of 1 × 1 × 1. In the encoder part, we doubled the channel number with the depth increasing to capture more global feature information, and we set the maximum of channels at 128 to speed up training. To prevent overfitting, we use well-known dropout techniques after each convolutional layer.[15] For the decoder, we use the double-channel strategies to construct the convolutional layers. All other layer details are the same as in the encoder.

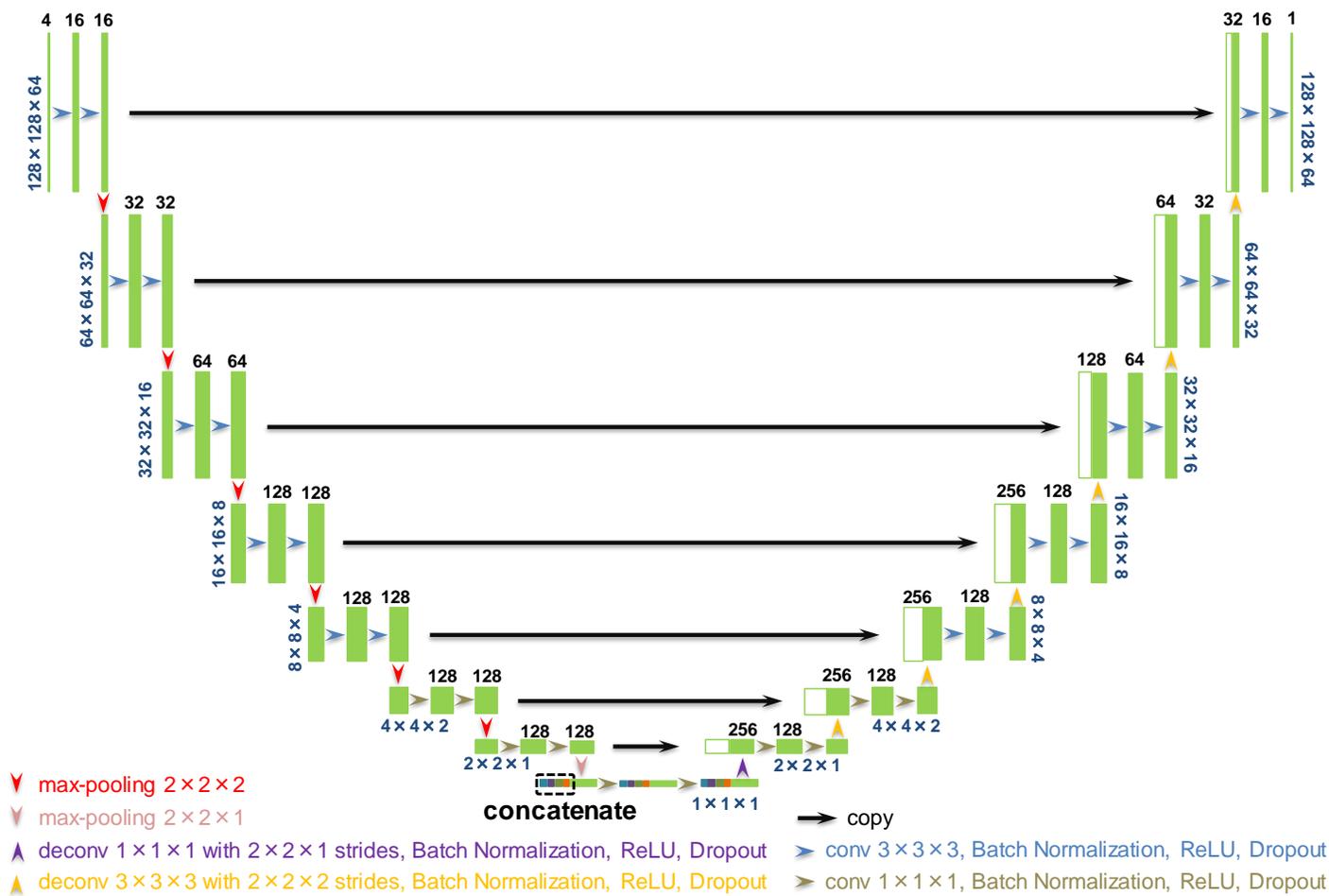

Figure 3. Modified 3D U-Net architecture with two inputs and one output. The green boxes denote multi-channel 3D feature maps, and each white box indicates a copied 3D feature map. The number at the top of the box represents the channel number for each feature map, and the map size is at the lower left corner of the box.

*2.2 Data Acquisition*

To demonstrate the feasibility of our proposed method, we used data from 97 patients with prostate cancer. We generated ten Pareto optimal IMRT treatment plans for each patient. Four critical structures—rectum, bladder, body, and PTV—for each patient were used in IMRT planning with a standard 7-beam protocol. We split the dataset into a training dataset of 77 patients and a testing dataset of 20 patients. The dimensions of contours and dose distributions were 128 × 128 × 64, and each DVH vector contained 32 elements, so the total number of DVH elements was 128, which is also the channel number of contours at the bottom of the encoder. All dose distributions were normalized by PTV mean dose to generate a uniform dataset to stabilize the training process.

Both model training and testing require physician-tuned desired DVHs $\mathcal{D}_i^k$ as model inputs. The corresponding dose distributions $\mathcal{Y}_i^k$ are required for training as labels and for testing for performance evaluation. Here, $\mathcal{Y}_i^k$ should be approximately Pareto optimal and should have DVHs close to $\mathcal{D}_i^k$ in the $l_1$-norm sense (Equation (4)). Therefore, we augmented the training dataset from $\{X_i, D_i^j, Y_i^j | i \in 1,\cdots,77; j \in 1,\cdots,10\}$ to $\{X_i, \mathcal{D}_i^k, \mathcal{Y}_i^k | i \in 1,\cdots,77; k \in 1,\cdots,K\}$ and the testing dataset from $\{X_i, D_i^j, Y_i^j | i \in 78,\cdots,97; j \in 1,\cdots,10\}$ to $\{X_i, \mathcal{D}_i^k, \mathcal{Y}_i^k | i \in 78,\cdots,97; k \in 1,\cdots,K'\}$. In this feasibility study, we generated $\mathcal{D}_i^k$ by randomly choosing a patient and a plan for patient $i$ from $\{X_i, D_i^j, Y_i^j | i \in 1,\cdots,77; j \in 1,\cdots,10\}$ for training and from $\{X_i, D_i^j, Y_i^j | i \in 78,\cdots,97; j \in 1,\cdots,10\}$ for testing to mimic the physician's desired DVH tuned on the fly during the training and testing process. The corresponding $D_i^j$ (and then $Y_i^j$) was selected through Equation (4). Here, $K$ is about 770 for training, and $K'$ is about 200 for testing.

*2.3 Implementation Details*

We adopted the Adam optimizer to minimize the loss function (5).[16] The batch size was 3, and the model was trained with 300 epochs. The learning rate decayed as the number of iterations increased, which can be defined as:

$$lr = \frac{lr_{initial}}{1+decay \times iterations} \qquad (6)$$

where the initial learning rate $lr_{initial}$ is set as 0.001, decay factor $decay$ is 0.002, and $iterations$ denotes the number of model weights updated. Since deeper layers can more easily be overfitted due to the greater number of weights, we gradually increased the dropout rate based on the following equation:

$$dropout = rate_{initial} \times \left(\frac{layer_{current}}{layer_{max}}\right) \qquad (7)$$

where the initial rate $rate_{initial}$ was set to 0.2 and the max layer $layer_{max}$ was equal to 7. In this work, we utilized a workstation equipped with 12 NVIDIA Tesla K80 GPUs to implement our network in Keras library with a TensorFlow[17] back end.

### 3. Results

Figure 4 shows the losses over epochs, where both the training loss and the validation loss follow a convergence trend, though validation loss has some oscillations. This implies that our model gradually reaches the optimal solution as the epochs increase.

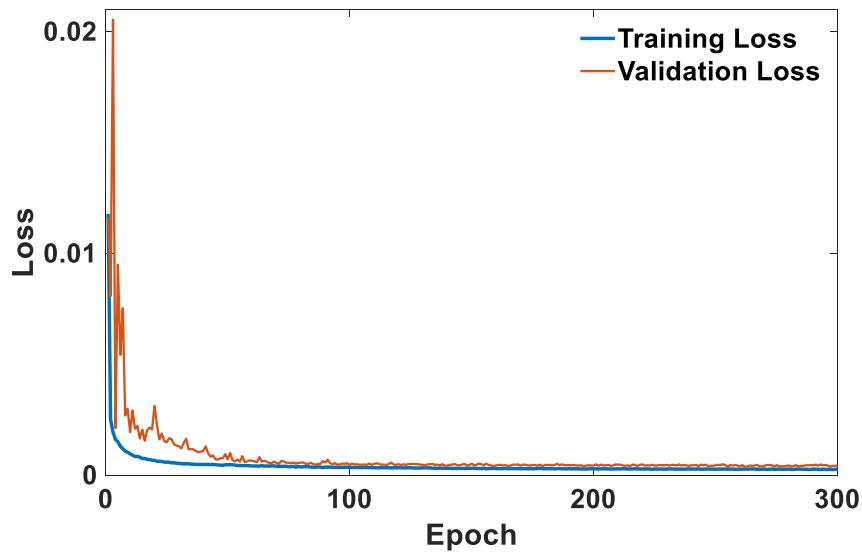

Figure 4. Loss values as a function of epochs in the training phase. Although the validation loss (orange line) has some oscillations, both the training loss (blue line) and the validation loss follow a convergence trend.

Figure 5 shows the individualized dose distributions predicted for a test patient, where each row represents a Pareto optimal plan of a different PTV/OAR trade-off for this patient. We can see that the predicted dose distributions are quite close to the corresponding true dose distributions for different PTV/OAR trade-offs. Although the dose distributions in the first and second rows are based on the same patient's anatomy, they are very different due to the different desired trade-offs. The predicted plan in the second row has much better rectum sparing with a slightly higher bladder dose, which might be preferable clinically.

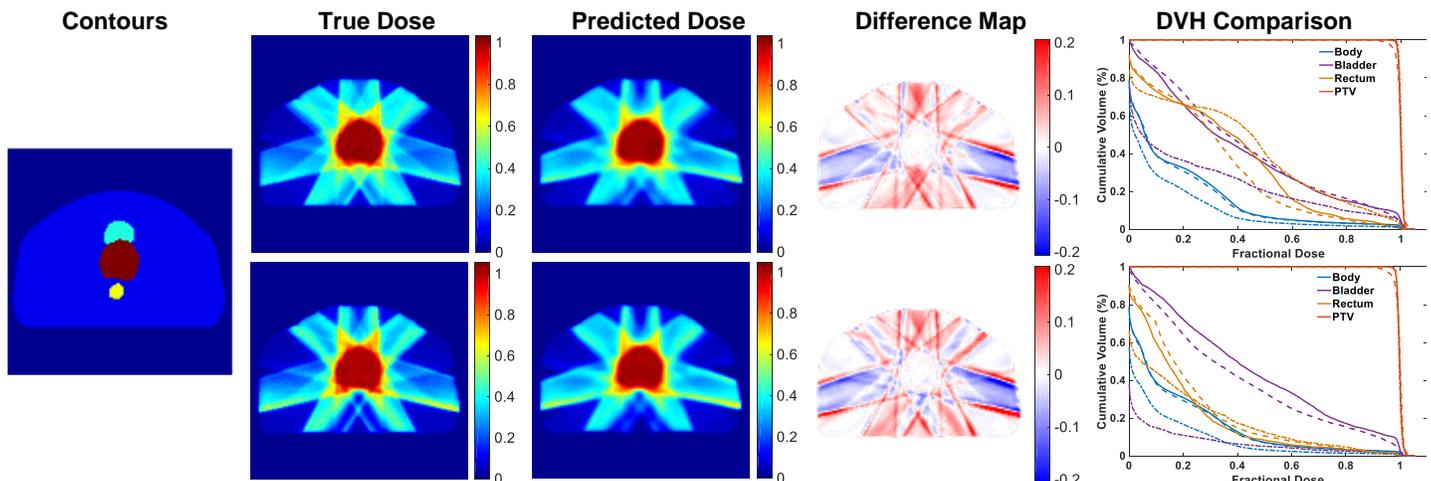

Figure 5. Individualized dose distributions predicted for a testing patient. From left column to right column: contours, true dose, predicted dose distribution, difference map (true – prediction), and DVH comparison between true (solid line) and predicted (dashed line) dose distributions. The input desired DVH curves are also shown in dotted lines. All dose values are normalized to the prescription dose. Each row represents a Pareto optimal plan of a different PTV/OAR trade-off for this patient.

For each OAR and PTV of the 20 test patients, we computed the differences between the predicted and the true mean and maximum doses. For all calculations, we subtracted the predicted dose from the true dose, then normalized the value to the prescription dose. The results are presented via violin plots in Figure 6 and via means and standard deviations in Table 1. We can see that the overall performance of the developed model is quite good. Specifically, the difference between the predicted and the true maximum OAR or PTV dose has both mean values and standard deviations within 2% of the prescription dose. The difference between the predicted and the true mean PTV doses is very small, with about 0.5% mean value and 0.2% standard deviation, because of the normalization to the prescription dose. The difference for body dose is also very small. However, the standard deviations of the differences between the predicted and the true mean rectum and bladder doses are quite large, close to 5%, though the mean values are still small (0.5% for rectum and 1.6% for bladder). The relatively large standard deviations for the mean dose differences in those two OARs may be explained by the small number of plans for each patient in the training and testing datasets ($N = 10$). This will be discussed in the next section.

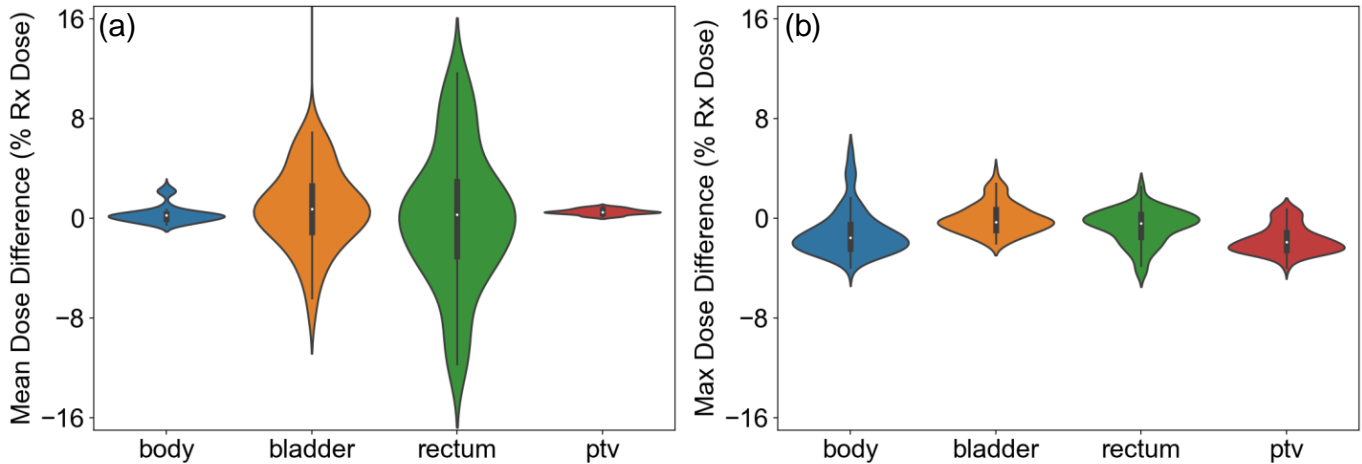

Figure 6. Violin plot of the differences, normalized to the prescription dose, between the predicted and true (a) mean and (b) maximum doses in each OAR and PTV for all test patients.

Table 1. Mean and maximum dose differences between the true and predicted doses in each OAR and PTV (*avg. ± std.*).

|         | Mean dose difference | Max dose difference |
|---------|----------------------|---------------------|
| Body    | 0.33 ± 0.73          | 1.21 ± 1.90         |
| Bladder | 1.63 ± 4.48          | 0.02 ± 1.17         |
| Rectum  | 0.54 ± 4.83          | 0.64 ± 1.46         |
| PTV     | 0.52 ± 0.23          | 1.83 ± 1.09         |

## 4. Discussion and Conclusions

Nguyen *et al.* discovered in 2017 that the relationship between the clinically optimal dose distribution and patient anatomy can be learned through supervised learning and then demonstrated that the trained DL model can predict the dose distribution given the PTV and OAR masks without going through the traditional treatment planning process [6-7]. Since then, many studies have exploited this idea [10-13]. In our clinic, we have implemented this work in routine clinical practice to assist physicians and treatment planners in producing treatment plans with higher efficiency and consistently higher quality. During the clinical implementation process, we realized that the predicted dose distributions represent a population average of the previously delivered treatment plans and often require further tuning to suit the treatment goal for a particular patient. When the physician tunes the predicted dose distribution to generate dose volume constraints as directives to guide the treatment planner, the tuned dose volume constraints may not correspond to a Pareto optimal plan and may not be achievable. Then, the clinical significance of using DL-based dose prediction to guide treatment planning will be greatly diminished.

To solve this problem, we developed another DL model that takes a desired DVH as input in addition to the patient's anatomy. We propose a new clinical workflow based on this DL model: 1) an anatomy-based DL model will be used to predict a population-averaged dose distribution and the corresponding DVH; 2) the physician or planner will tune the predicted DVH to reflect the individualized planning goals for a particular patient; and 3) the proposed DL model will be used to predict the refined dose distribution and DVH using the patient anatomy and the tuned DVH as inputs. Steps 2) and 3) can be repeated until the desired dose distribution and DVH are achieved.

This paper presents a feasibility study of the proposed DL model. We used 77 patients treated with IMRT for prostate cancer for model training and 20 patients for testing. Each patient has 10 plans corresponding to different PTV/OAR trade-offs. The results shown in Figures 5 and 6 and Table 1 demonstrate the feasibility of the proposed model. The relatively large standard deviations (~5% of the prescription dose) for the mean dose differences in rectum and bladder are likely due to the small number of plans for each patient in the training and testing datasets. With only 10 plans, the Pareto surface can only be covered with a very low resolution, which led to the sub-optimal model performance in the testing phase. When the proposed method is used clinically, much more Pareto optimal plans should be generated for each training patient. This could be very time-consuming, but it could be done automatically by scripting the planning systems overnight when the systems are not in use clinically.

Although we have demonstrated the feasibility of the proposed method, a much more comprehensive and clinically realistic performance evaluation study should be conducted before it can be used clinically. Ideally, the trained DL model needs to be integrated with other clinical software and tested by physicians and treatment planners on real patient cases. An important metric for evaluation should be the efficiency gain with this method by comparing it with the traditional methods of iterative communication between physicians and planners.

In conclusion, we have developed and demonstrated the feasibility of a DL model that can predict individualized and approximately Pareto optimal 3D dose distributions by using the desired DVH as input in addition to the patient's anatomy. This model can facilitate a new clinical workflow to guide treatment planning based on DL-based dose prediction.

**Acknowledgements**

This work was supported by an NIH R01 grant (1R01CA237269-01). We would like to thank Dr. Jonathan Feinberg for editing the manuscript.